 \documentclass[12pt]{article}
 \usepackage[cp1251]{inputenc}
 \usepackage[english, russian]{babel}
 \usepackage{amsmath, latexsym, amssymb, bm, array, graphics, amsfonts}


\usepackage[dvips]{graphicx}

\begin{document}
\thispagestyle{empty}
\large
\renewcommand{\abstractname}{Abstract}
\renewcommand{\refname}{\begin{center}
 REFERENCES\end{center}}
\makeatother

\begin{center}
\bf Analytical solution of problem about moderately strong
evaporation (condensation) for one-dimensional
kinetic equation
\end{center} \medskip

\begin{center}
  \bf
  A. V. Latyshev\footnote{$avlatyshev@mail.ru$} and
  A. A. Yushkanov\footnote{$yushkanov@inbox.ru$}
\end{center}\medskip

\begin{center}
{\it Faculty of Physics and Mathematics,\\ Moscow State Regional
University, 105005,\\ Moscow, Radio str., 10A}
\end{center}\medskip

\begin{abstract}
For one-dimensional linear kinetic equations analytical
solutions of prob\-lems about moderately strong evaporation
(condensation), when frequency of col\-li\-sions of
molecules is constant, are received . The equation
and distri\-bution function  are linearize concerning the absolute
Maxwellian, given far from a wall.
Quantities of  of temperature and concentration jumps are found.
Distributions of concentration, mass velocity and temperature
are constructed.


\medskip


\end{abstract}

\begin{center}
\bf   Введение
\end{center}
History of this problem in linear statement and with application
one-dimensional collisional integral is stated in works \cite{1,2}.
In these works attempts of the exact solution of problem
about "strong"\, liquid evapo\-ration in vacuum  have been undertaken.
Thus linearization was spent concerning equilibrium
distribution function, given far from an eva\-po\-ration surface.
It has allowed to consider influence of movement of gas from a wall
on behaviour of gas in Knudsen layer, remaining thus in frameworks
linear approach. Velocity of the expiration of gas (and others
parametres) were included into distribution function by nonlinear method.
It is possible to name the similar approach "quasilinear"\,. Des\-pite the
model solution and problem, it allows to describe correctly a number
of the basic qualitative characteristics of evaporation.
Them concerns first of all separation Mach number equal to unit
($ {\bf M}=1$).

In work \cite{1} for the solution of problem the method of
resolvent has been used. And in \cite{2,3} a method of
boundary value problem from theory of functions of complex variable
has been used. In \cite{4} with attraction of methods of the
functional analysis was resolvability of a problem in one
interval
($0 <U <\sqrt {3/2} $) is shown and
unsolvability in other $U> \sqrt {3/2} $, $U $ is the dimensionless
velocity of evaporation.
In work \cite{5} approximation $F_N $-method has been used.
In the monography \cite{6} (Chapter III, \S 4)
this problem was studied in abstract statement.

Most nearly to the exact solution of this problem  in a case of
eva\-po\-ra\-tion Siewert and Thomas  have approached \cite{2}, and in a case
condensation --- Cercignani and Frezzotti \cite{3}. However in these
works \cite{2,3} authors could not receive exact relations
between quantities of temperature jump, jump of density and
velocity of evaporation (concentration).

The analytical solution of this problem has been finished in
our work \cite{77}.

In works \cite{8}--\cite{11} the linear one-dimensional kinetic
equation with collisional integral of BGK (Bhatnagar, Gross and
Krook) and frequency of collisions, affine depending on the module
of molecular velocity was considered.

The considered physical problem consists in the solution of the boun\-dary
problems for the modelling kinetic equation. Required
physical quantities contain in boundary conditions.
Substitution (Case' ansatz)
$$
h_\eta(x,\mu)=\exp(-\dfrac{x}{\eta+U})\Phi(\eta,\mu)
$$
at once reduces the kinetic equation to the characteristic
equation.
From the solution of the characteristic equation are found
eigen functions in space of generalized functions.

Further the structure of discrete and continuous spectra of the
charac\-te\-ris\-tic equations is investigated.
Further the theorem about expansion of solution
of boundary problem on generalized eigen functions  is proved.
The proof is reduced to the solution of singular integral
equations with Cauchy kernel.
This equation is reduced to the solution of the Riemann boundary
value problem on semiaxis $ [-U, + \infty) $. After the solution
of corresponding homogeneous boundary value problem is found the general
solution of a non-homogeneous boundary value problem. Unknown coef\-fi\-cients
(physical quantities) of solution expansion from conditions of
resolvability of boundary value problem  are found.

\begin{center}
  {\bf 1. Statement of  problem and  basic equations}
\end{center}

Let us consider evaporation (condensation) of a liquid from the flat
surfaces $x=0$ in the vacuum occupying half-space $x> 0$.
We take one-dimensional BGK--equation
$$
\zeta\dfrac{\partial f}{\partial
x}=\nu[\Phi(x,\zeta)-f(x,\zeta)],
\eqno{(1.1)}
$$
where $f(x,\zeta)$ is the distribution function, $\zeta$ is the
molecular velocity in the axes direction $x$, $\nu$ is the
collisional frequency, $\Phi(x,\zeta)$ is the local Maxwellian,
$$
\Phi(x,\zeta)=\dfrac{\rho(x)}{\sqrt{2\pi RT(x)}}\exp\Big\{
-\dfrac{[\zeta-v(x)]^2}{2RT(x)}\Big\}.
$$

Here
$$
\rho(x)=\int\limits_{-\infty}^{\infty}f(x,\zeta)d\zeta,
$$
$$
v(x)=\dfrac{1}{\rho(x)}\int\limits_{-\infty}^{\infty}\zeta f(x,\zeta)
d\zeta,
$$
$$
T(x)=\dfrac{1}{R\rho(x)}\int\limits_{-\infty}^{\infty}
[\zeta-v(x)]^2f(x,\zeta)d\zeta.
$$

Jumps of temperature and density are required to be found
$$
\varepsilon_T=\dfrac{T_s-T_\infty}{T_\infty},\qquad
\varepsilon_\rho=\dfrac{\rho_s-\rho_\infty}{\rho_\infty},
$$
where $T_s,\;\rho_s$ are temperature and density of gas
directly nearby at a wall.

Also density distribution $\rho(x)$,
mass velocity $v(x)$ and temperatures at $x> 0$ are required to be found.

We believe that molecules are reflected purely diffusively from
a wall. It means, that molecules are reflected from a wall with
Maxwell distribution, i.e.
$$
f(x=0,\zeta)=f_s(\zeta),\qquad  \zeta>0,
$$
where
$$
f_s(\zeta)=\dfrac{\rho_s}{\sqrt{2\pi RT_s}}\exp\Big(-\dfrac{\zeta^2}
{2RT_s}\Big).
$$

Let us assume, that far from a surface steam condition is described
by equilibrium distribution characterized by the constant
velocity of evaporation (condensation) $v_\infty $, density
$ \rho_\infty $ and temperature $T_\infty $, i.e.
$$
\Phi(\infty,\zeta)\equiv f_\infty(\zeta)=
\dfrac{\rho_\infty}{\sqrt{2\pi RT_\infty}}\exp\Big\{
-\dfrac{[\zeta-v_\infty]^2}{2RT_\infty}\Big\}.
$$

Following \cite{1}, we will be linearize function of distribution and local
Maxwellian concerning $f_\infty (\zeta) $. Entering the shift
variable $c =\zeta-v_\infty $, we will write
$$
f(x,\zeta)=f_\infty(c)[1+h(x,c)],
\eqno{(1.2)}
$$
where
$$
f_\infty(\zeta)=
\dfrac{\rho_\infty}{\sqrt{2\pi RT_\infty}}\exp\Big\{
-\dfrac{c^2}{2RT_\infty}\Big\}.
$$

We put in linear approximation
$$
\rho(x)=\rho_\infty+\delta\rho(x),
$$
$$
T(x)=T_\infty+\delta T(x),
$$
$$
v(x)=v_\infty+\delta v(x).
$$

Let us pass to dimensionless variables: to dimensionless coordinate
$x_1$, dimensionless molecular speed $ \mu $, dimensionless
mass velocity of gas $U(x)$
$$
x_1=\dfrac{\nu x}{\sqrt{2RT_\infty}},\qquad
\mu=\dfrac{c}{\sqrt{2RT_\infty}},\qquad
U(x)=\dfrac{v(x)}{\sqrt{2RT_\infty}}.
$$

Further dimensionless coordinate
$x_1$ we will designate again through $x$.

Let us enter also the dimensionless given velocity of
evaporation (condensation)
$$
U=U_\infty=U(\infty)=\dfrac{v_\infty}{\sqrt{2RT_\infty}}.
$$

By means of these designations we receive, that in linear approach
$$
\Phi(x,\zeta)-f(x,\zeta)=$$$$=
f_\infty(\mu)\Big[\dfrac{\delta \rho (x)}{\rho_\infty}+2\mu \delta U(x)+
\Big(\mu^2-\dfrac{1}{2}\Big)\dfrac{\delta T(x)}{T_\infty}
-h(x,\mu)\Big].
$$

Let us consider distributions of density, mass velocity and
temperature also we will express their relative changes with help
of function $h(x,\zeta)$.

For density we have
$$
\dfrac{\rho(x)}{\rho_\infty}=1+\dfrac{\delta \rho(x)}{\rho_\infty},
\qquad
\dfrac{\delta \rho(x)}{\rho_\infty}=\dfrac{1}{\sqrt{\pi}}
\int\limits_{-\infty}^{\infty}e^{-\mu^2}h(x,\mu)d\mu.
\eqno{(1.3)}
$$

For mass velocity we have
$$
U(x)=U+\dfrac{1}{\sqrt{\pi}}
\int\limits_{-\infty}^{\infty}e^{-\mu^2}\mu h(x,\mu)d\mu.
\eqno{(1.4)}
$$

For temperature distribution we have
$$
T(x)=\dfrac{1}{R\rho(x)}\int\limits_{-\infty}^{\infty}[c-\delta
v(x)]^2f_\infty(c)[1+h(x,c)]dc=
$$
$$
=\dfrac{1}{R\rho(x)}\int\limits_{-\infty}^{\infty}c^2f_\infty(c)dc+
\dfrac{1}{R\rho(x)}\int\limits_{-\infty}^{\infty}c^2f_\infty(c)h(x,c)]dc.
$$

From here, passing to dimensionless velocity of integration,
we receive
$$
\dfrac{T(x)}{T_\infty}=\dfrac{\rho_\infty}{\rho(x)}+
2\dfrac{\rho_\infty}{\rho(x)}\dfrac{1}{\sqrt{\pi}}
\int\limits_{-\infty }^{\infty}e^{-\mu^2}\mu^2h(x,\mu)d\mu.
$$

Noticing that
$$
\dfrac{\rho_\infty}{\rho(x)}=1-\dfrac{\delta
\rho(x)}{\rho_\infty},
$$
From the previous we receive
$$
\dfrac{T(x)}{T_\infty}=\dfrac{2}{\sqrt{\pi}}
\int\limits_{-\infty }^{\infty}e^{-\mu^2}
\Big(\mu^2-\dfrac{1}{2}\Big)h(x,\mu)d\mu.
\eqno{(1.5)}
$$

Now we can definitively formulate the one-dimensional
linear kinetic equation with integral of collisions
in the form of BGK
$$
(\mu+U)\dfrac{\partial h}{\partial x}+h(x,\mu)=\dfrac{1}{\sqrt{\pi}}
\int\limits_{-\infty}^{\infty}e^{-\mu'^2}q(\mu,\mu')h(x,\mu')d\mu'.
\eqno{(1.6)}
$$

Here $q(\mu,\mu')$ is the kernel (or indicatrix) of equation,
$$
q(\mu,\mu')=1+2\mu\mu'+2\Big(\mu^2-\dfrac{1}{2}\Big)
\Big(\mu'^2-\dfrac{1}{2}\Big).
$$

Let us understand with boundary conditions. As distribution function
far from a wall passes in Maxwell distribution function,
given far from a wall
$$
 \lim\limits_{x\to +\infty}f(x,\zeta)=f_\infty(\zeta),
$$
from here follows, that for function $h(x,\zeta) $ at once follows
boundary condition far from a wall
$$
h(+\infty,\mu)=0.
\eqno{(1.7)}
$$

The condition diffusion reflexion  of molecules from a wall means, that
$$
f_\infty(c)[1+h(0,c)]=f_s(c),\qquad c>-v_\infty.
$$
From here we receive, that
$$
h(0,c)=\dfrac{\rho_s}{\rho_\infty}\sqrt{\dfrac{T_\infty}{T_s}}
\exp\Big[-\dfrac{(c+v_\infty)^2}{2RT_s}+\dfrac{c^2}{2RT_\infty}\Big]
$$

In linear approximation we have
$$
-\dfrac{(c+v_\infty)^2}{2RT_s}+\dfrac{c^2}{2RT_\infty}=
-\dfrac{(\mu+U)^2}{1+\varepsilon_T}+\mu^2=-2U\mu+\varepsilon_T\mu^2.
$$

Hence, from here we receive the second boundary condition
$$
h(0,\mu)=\varepsilon_\rho-2U\mu+\varepsilon_T\Big(\mu^2-\dfrac{1}{2}\Big),
\qquad \mu>-U.
\eqno{(1.8)}
$$

So, the boundary problem about moderately strong evaporation
(con\-densation) for one-dimensional gas consists in finding of such
solution of the equation (1.6) which satisfies to boundary
conditions (1.7) and (1.8).

\begin{center}
  \bf 2. Separation of variables. Dispersion function. General
solution of kinetic equation
\end{center}

Substitution (ansatz of Case)
$$
h_\eta(x,\mu)=\exp\Big(-\dfrac{x}{U+\eta}\Big)\Phi(\eta,\mu)
\eqno{(2.1)}
$$
at once reduces the equation (1.6) to the characteristic
$$
(\eta-\mu)\Phi(\eta,\mu)=
$$
$$
=\dfrac{1}{\sqrt{\pi}}(\eta+U)\Big[n_{(0)}(\eta)+
2\mu n_1(\eta)+2\Big(\mu^2-\dfrac{1}{2}\Big)\Big(n_2(\eta)-
\dfrac{1}{2}n_0(\eta)\Big)\Big].
\eqno{(2.2)}
$$

Here the designation is entered
$$
n_k(\eta)=\int\limits_{-\infty}^{\infty}e^{-\mu^2}\Phi(\eta,\mu)\mu^kd\mu,
\qquad k=0,1,2.
$$

Multiplying the equation (1.6) on $ \mu^ke^{-\mu^2} \; (k=0,1) $ and
integrating on all real axis, we receive two equations
$$
n_1(\eta)=-Un_0(\eta)
$$
and
$$
n_2(\eta)=-Un_1(\eta)=U^2n_0(\eta).
$$

By means of two last equalities the equation (2.2) will be
rewritten in the form
$$
(\eta-\mu)\Phi(\eta,\mu)=\dfrac{1}{\sqrt{\pi}}(\eta+U)q(-U,\mu)n_0(\eta),
\eqno{(2.3)}
$$
where
$$
q(-U,\mu)=1-2U\mu+2\Big(U^2-\dfrac{1}{2}\Big)\Big(\mu^2-\dfrac{1}{2}\Big).
$$

Further we will accept the following condition of normalization
$$
n_0(\eta)\equiv\int\limits_{-\infty}^{\infty}e^{-\mu^2}\Phi(\eta,\mu)d\mu
\equiv 1.
\eqno{(2.4)}
$$

Now the characteristic equation
$$
(\eta-\mu)\Phi(\eta,\mu)=\dfrac{1}{\sqrt{\pi}}(\eta+U)q(-U,\mu)
$$
has in space of generalized functions the following solution
$$
\Phi(\eta,\mu)=\dfrac{1}{\sqrt{\pi}}(\eta+U)q(-U,\mu)P\dfrac{1}{\eta-\mu}+
g(\eta)\delta(\eta-\mu),
\eqno{(2.5)}
$$
where
$$
 \eta,\mu\in (-\infty,+\infty).
$$
Here $Px^{-1}$ means distribution --- a principal value
of integral at integ\-ra\-tion of expression $x^{-1}$, $ \delta(x) $
is the Dirac delta-function.

Substituting expression (2.5) in the condition of
normalization (2.4), we find, that
$$
g(\eta)=e^{\eta^2}\lambda(\eta),
$$
where $\lambda(z)$ is the dispersion function,
$$
\lambda(z)=1+\dfrac{z+U}{\sqrt{\pi}}\int\limits_{-\infty}^{\infty}
e^{-\mu^2}\dfrac{q(-U,\mu)}{\mu-z}d\mu.
\eqno{(2.6)}
$$

Thus, eigen functions of the characteristic equation,
corresponding to continuous spectrum, look like
$$
\Phi(\eta,\mu)=\dfrac{1}{\sqrt{\pi}}(\eta+U)q(-U,\mu)P\dfrac{1}{\eta-\mu}+
e^{\eta^2}\lambda(\eta)\delta(\eta-\mu).
\eqno{(2.5')}
$$

According to (2.1), eigen solution of equation (1.6) decreasing
far from a wall and corresponding to continuous spectrum, look like
$$
h_\eta(x,\mu)=\exp\Big(-\dfrac{x}{\eta+U}\Big)\Phi(\eta,\mu),\quad
\eta>-U, \quad -\infty<\mu<+\infty.
$$

Expression (2.6) for dispersion function it is possible to express thro\-ugh
dispersion function of plasma
$$
\lambda_C(z)=1+\dfrac{z}{\sqrt{\pi}}\int\limits_{-\infty}^{\infty}
\dfrac{e^{-\mu^2}d\mu}{\mu-z}=\dfrac{1}{\sqrt{\pi}}
\int\limits_{-\infty}^{\infty}
\dfrac{\mu e^{-\mu^2}d\mu}{\mu-z}.
$$

Substituting an explicit form of function $q(-U, \mu) $ in (2.6),
we receive, that
$$
\lambda(z)=1+(z+U)\bigg\{\Big[2\Big(U^2-\dfrac{1}{2}\Big)z-2U\Big]
\lambda_C(z)+\Big(\dfrac{3}{2}-U^2\Big)t(z)\bigg\}.
$$

According to formulas of Sokhotsky, for boundary values
dispersion function we have
$$
\lambda^{\pm}(\mu)=\lambda(\mu)+i\sqrt{\pi}(\mu+U)e^{-\mu^2}q(-U,\mu),\quad
-\infty<\mu<+\infty.
$$

Here
$$
\lambda(\mu)=1+\dfrac{\mu+U}{\sqrt{\pi}}\int\limits_{-\infty}^{\infty}
e^{-\mu'^2}\dfrac{q(-U,\mu')}{\mu'-\mu}d\mu',
$$
and integral in this equality is understood as singular in sense
of principal value on Cauchy.

From formulas of Sokhotsky we receive, that
$$
\lambda^+(\mu)-\lambda^-(\mu)=2\sqrt{\pi}i(\mu+U)e^{-\mu^2}q(-U,\mu),
$$
$$
\dfrac{\lambda^+(\mu)+\lambda^-(\mu)}{2}=\lambda(\mu), \quad
-\infty<\mu<+\infty.
$$

By means of an argument principle \cite{26} it is possible to show, that
dispersion function has no complex zero in the finite
parts of complex plane.

Let us expand dispersion function in Laurent series in a vicinity
infinitely remote point
$$
\lambda(z)=-U\Big(U^2-\dfrac{3}{2}\Big)\dfrac{1}{z^3}-$$$$-
\dfrac{3}{2}\Big(U^2-\dfrac{1}{2}\Big)\dfrac{1}{z^4}-
3U\Big(U^2-\dfrac{3}{2}\Big)\dfrac{1}{z^5}+\cdots, \quad z\to \infty.
\eqno{(2.7)}
$$

From expansion (2.7) it is visible, that infinitely remote point
is zero of the third order, if $U\neq 0$ and $U^2\neq
\dfrac{3}{2}$, and zero of the fourth order otherwise, i.e.
if $U=0$ or $U^2 =\dfrac{3}{2}$. This last case was
investigated in work \cite{78}.

To infinitely remote point, as to a multiple point of the discrete
spectrum, there correspond following discrete modes (solutions)
$$
h_0(x,\mu)=1,
$$
$$
h_1(x,\mu)=\mu,
$$
$$
h_2(x,\mu)=\mu^2,
$$
and
$$
h_3(x,\mu)=\Big(\mu^2-\dfrac{3}{2}\Big)(x-{\rm sign\;}\mu),
$$
if $U=0$ or $U^2=\dfrac{3}{2}$.

\begin{center}
  \bf 3.  Analytical solution of the problem about moderately strong
evaporation (condensation)
\end{center}

Having eigen solutions corresponding to continuous and discrete
spec\-tra, we construct the general solution of the equation
(1.6) in the form of integral on the continuous spectrum and the
linear combination of

\begin{figure}[t]\center
\includegraphics[width=16.0cm, height=10cm]{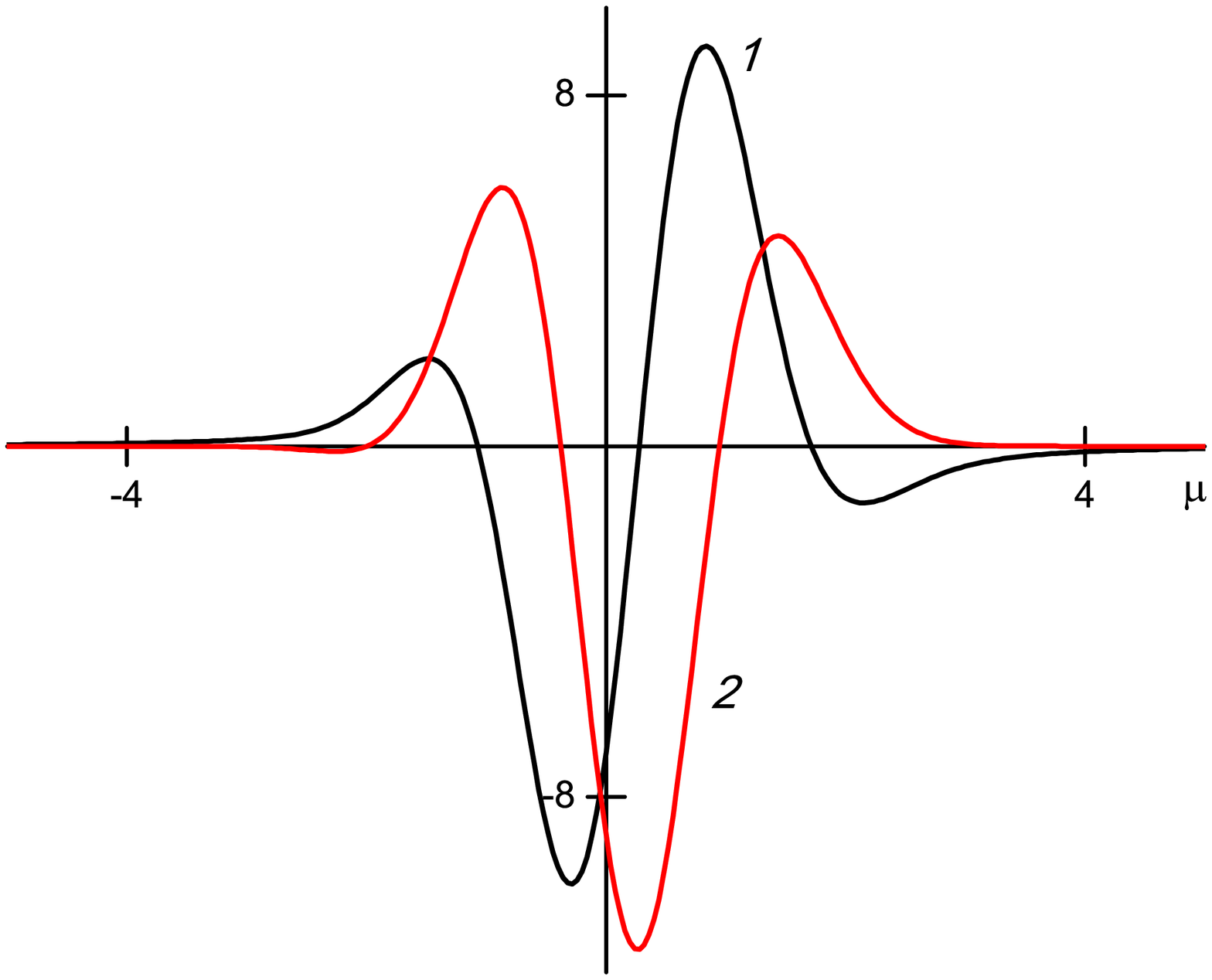}
{{\bf Fig. 1.} Real (curve 1) and imaginary (curve 2)
parts of dispersion function, $U=2$.}
\includegraphics[width=17.0cm, height=10cm]{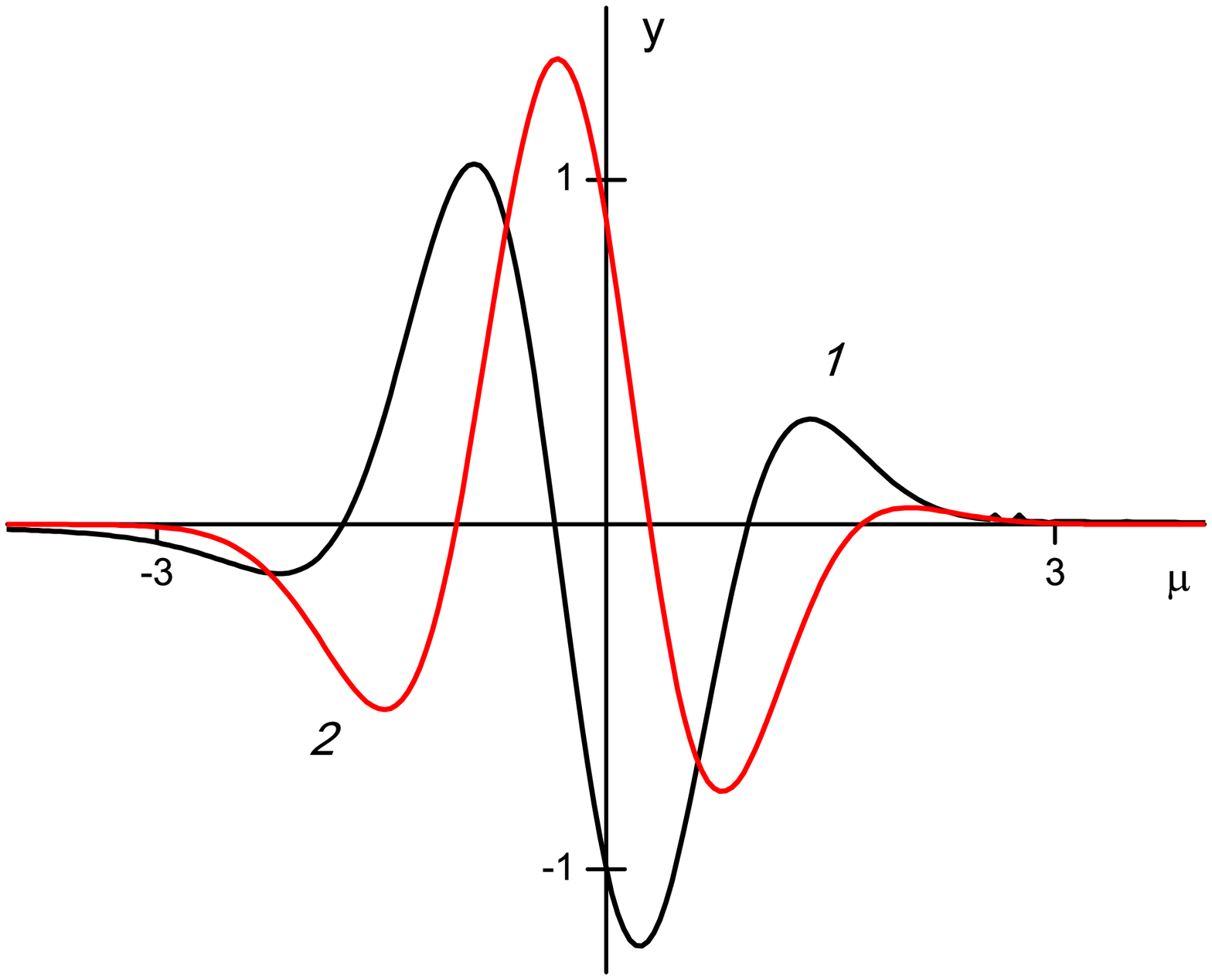}
{{\bf Fig. 2.} Real (curve 1) and imaginary (curve 2)
parts of dispersion function, $U=1$.}
\end{figure}
\clearpage
discrete eigen solutions
$$
h(x,\mu)=A_0+A_1\mu+A_2\mu^2+\int\limits_{-\infty}^{+\infty}
e^{-x/(\eta+U)}\Phi(\eta,\mu)a(\eta)d\eta.
\eqno{(3.1)}
$$

Constants $A_0, A_1, A_2$ and function $a (\eta) $ are
coefficients of expansion (3.1) corresponding  accordingly
to discrete and continuous spectra. These coefficients
are defined from boundary conditions (1.7) and (1.8).

Substituting expansion (3.1) in the boundary condition (1.7), we receive,
that $A_0=A_1=A_2=0$ and $a(\eta) =0$ at $ \eta <-U $. Hence,
expansion (3.1) becomes simpler
$$
h(x,\mu)=\int\limits_{-U}^{+\infty}
e^{-x/(\eta+U)}\Phi(\eta,\mu)a(\eta)d\eta.
\eqno{(3.2)}
$$

Let us substitute now expansion (3.2) in the boundary condition (1.8).
We receive the following integral equation
$$
\varepsilon_\rho-2U\mu+\varepsilon_T\Big(\mu^2-\dfrac{1}{2}\Big)=
\int\limits_{-U}^{+\infty}\Phi(\eta,\mu)a(\eta)d\eta,
\qquad \mu>-U.
\eqno{(3.3)}
$$

Substituting in (3.3) eigen functions of the continuous spectrum
(2.5'), we come to the singular integral equation with Cauchy kernel
$$
h(0,\mu)=q(-U,\mu)\dfrac{1}{\sqrt{\pi}}\int\limits_{-U}^{\infty}
\dfrac{(\eta+U)a(\eta)}{\eta-\mu}d\eta+e^{\mu^2}\lambda(\mu)a(\mu),
\eqno{(3.4)}
$$
where $\mu>-U,$
$$
h(0,\mu)=\varepsilon_\rho-2U\mu+\varepsilon_T\Big(\mu^2-\dfrac{1}{2}\Big).
$$

Let us enter auxiliary function
$$
N(z)=\dfrac{1}{\sqrt{\pi}}\int\limits_{-U}^{\infty}
\dfrac{(\eta+U)a(\eta)}{\eta-z}d\eta.
\eqno{(3.5)}
$$

This function is analytic in the complex plane $\mathbb{C}$ with
cut $ \mathbb{R}_U = [-U, + \infty] $. Its boundary values
from above and from below on the cut satisfy to formulas of Sokhotsky
$$
N^+(\mu)-N^-(\mu)=2\sqrt{\pi}i(\eta+U)a(\eta),
$$
$$
\dfrac{N^+(\mu)+N^-(\mu)}{2}=N(\mu),
$$
where
$$
N(\mu)=
\dfrac{1}{\sqrt{\pi}}\int\limits_{-U}^{\infty}
\dfrac{(\eta+U)a(\eta)}{\eta-\mu}d\eta,
$$
and last integral is understood in sense of a principal value.

By means of boundary values of auxiliary function $N(z)$ and
disper\-si\-on function $\lambda(z)$ we will reduce the singular
integral equation (3.4) to  nonhomogeneous boundary value problem
$$
\lambda^+(\mu)[q(-U,\mu)N^+(\mu)-h(0,\mu)]=$$$$=
\lambda^-(\mu)[q(-U,\mu)N^-(\mu)-h(0,\mu)],\quad \mu>-U.
\eqno{(3.6)}
$$

For the solution of a nonhomogeneous boundary value problem (3.6)
we will solve at first corresponding homogeneous Riemann boundary
value problem
$$
\dfrac{X^+(\mu)}{X^-(\mu)}=\dfrac{\lambda^+(\mu)}{\lambda^-(\mu)},\quad
\mu>-U.
\eqno{(3.7)}
$$

In the problem (3.7) unknown function $X(z)$ is analytic in cut
comp\-lex plane $ \mathbb{C} $ with
cut $ \mathbb{R}_U = [-U, + \infty] $.
The solution of the problem (3.7)
essentially depends on quantity and  sign of parametre $U $, from
which dispersion function  depends and which defines
semiaxis on which the boundaty value problem is given.

We notice that
$$
|\lambda^+(\mu)|=|\lambda^-(\mu)|,\qquad \lambda^+(\mu)=
\overline{\lambda^-(\mu)},\quad -\infty<\mu<+\infty.
$$

Let us enter the angle $ \theta(\mu)=\arg\lambda^+(\mu) $
is the principal value of argument fixed in the point
$\mu =-U $ by the condition $ \theta (-U) =0$ (see

\begin{figure}[t]\center
\includegraphics[width=16.0cm, height=10cm]{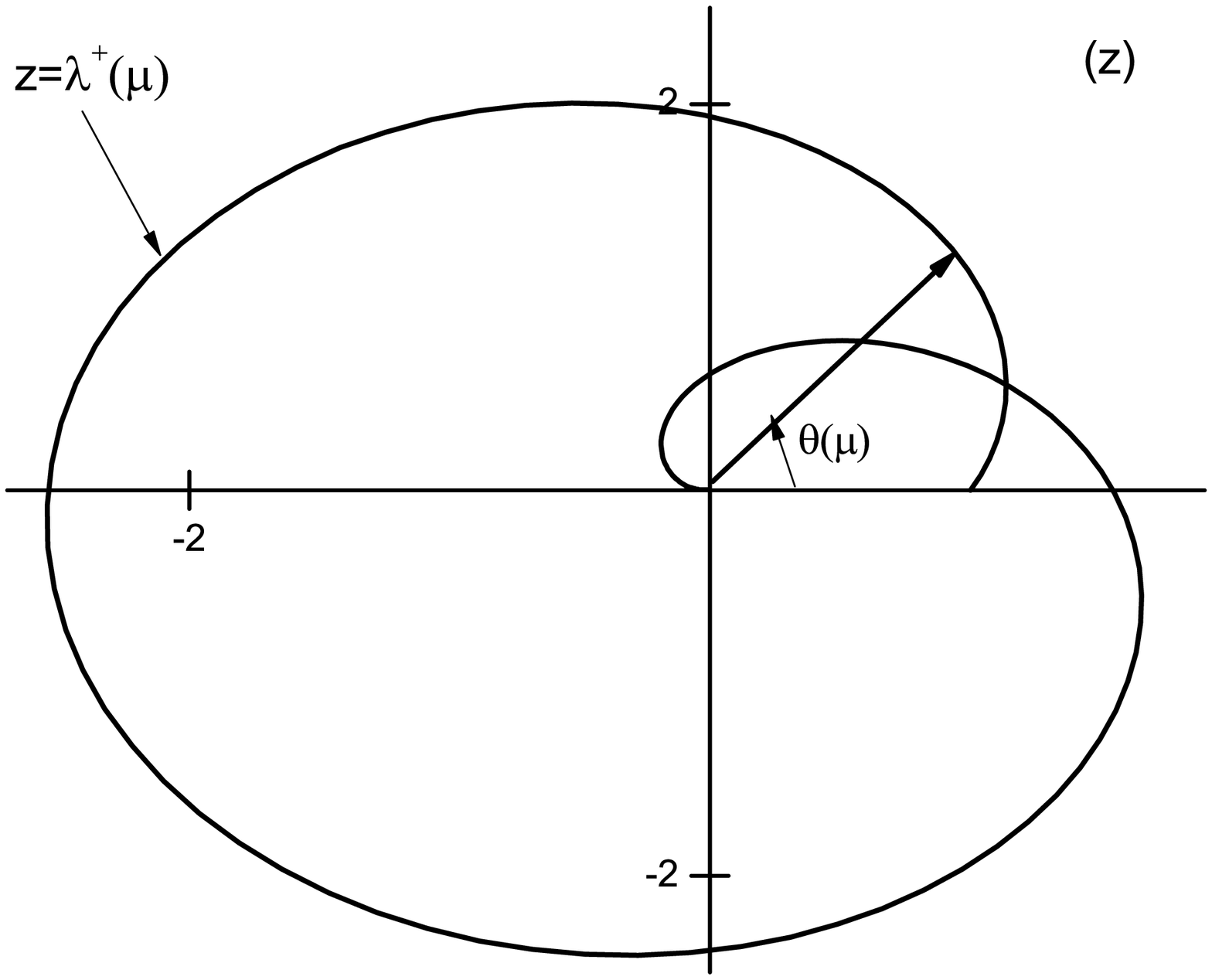}
{{\bf Fig. 3.}
The curve $z=\lambda^+(\mu), -(\sqrt{3/2}+0.1)\leqslant\mu\leqslant +\infty$,
the case $U=\sqrt{3/2}+0.1$. Incrementation of angle $\theta(\mu)$
equalsо $3\pi$ on semiaxis
$ -(\sqrt{3/2}+0.1)\leqslant \mu \leqslant+\infty$.}
\includegraphics[width=17.0cm, height=10cm]{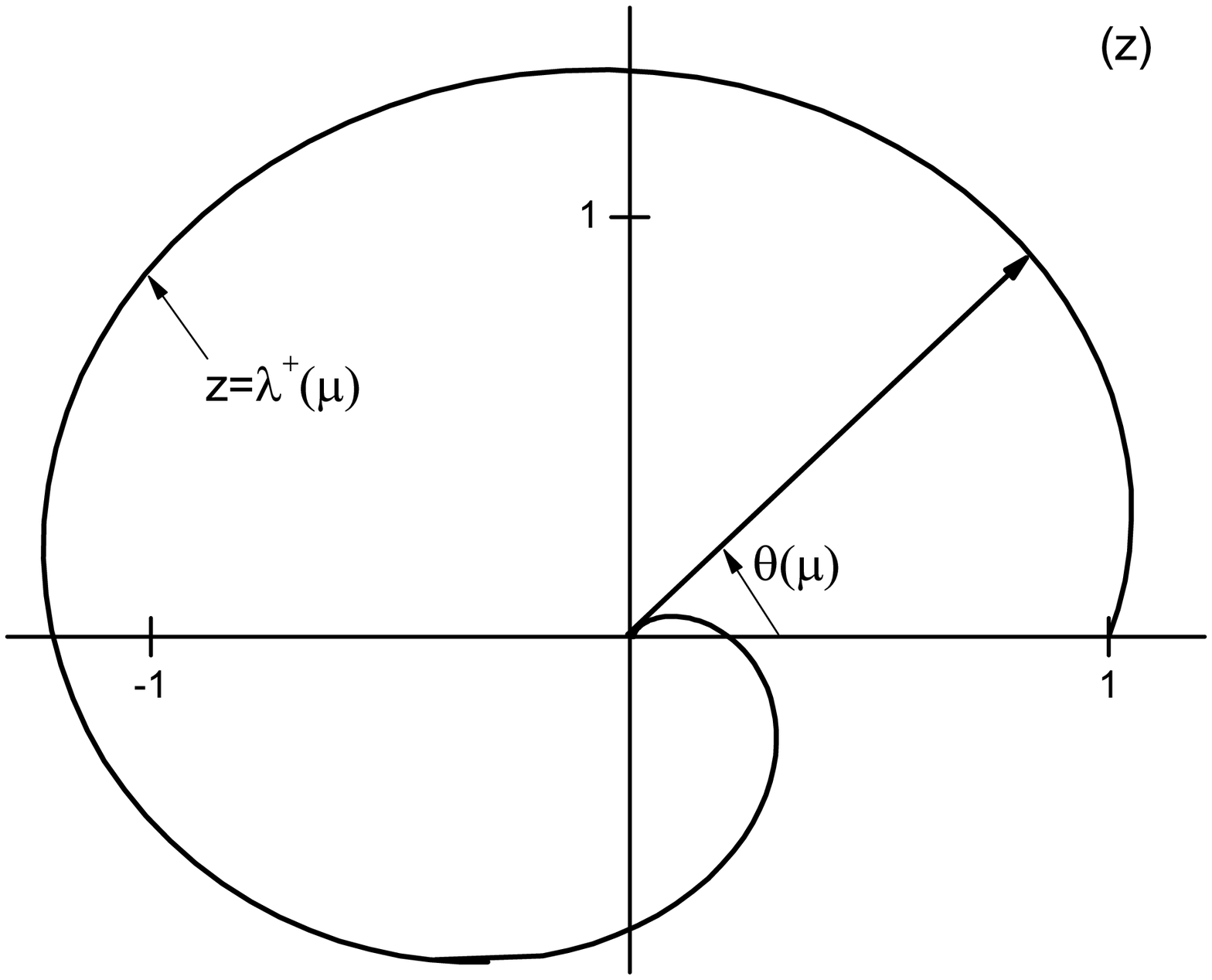}
{{\bf Fig. 4.} The curve
$z=\lambda^+(\mu), -1\leqslant \mu \leqslant+\infty$,
the case  $U=1$. Incrementation of angle $\theta(\mu)$ equals $2\pi$
on semiaxis $-1\leqslant \mu \leqslant+\infty$.}
\end{figure}
\clearpage
Figs. 3 and 4). It is easy to see, that
coefficient of the boundary va\-lue problem (3.7)
$$
G(\mu)=\dfrac{\lambda^+(\mu)}{\lambda^-(\mu)}
$$
is equal
$$
G(\mu)=\dfrac{\lambda^+(\mu)}{\lambda^-(\mu)}=
\dfrac{\lambda(\mu)+is(\mu)}{\lambda(\mu)-is(\mu)}=e^{2i \theta(\mu)}.
$$

Here
$$
s(\mu)=\sqrt{\pi}e^{-\mu^2}(\mu+U)q(-U,\mu),
$$
$$
\lambda(\mu)=1+(\mu+U)\bigg\{\lambda_C(\mu)
\Big[2\Big(U^2-\dfrac{1}{2}\Big)\mu-2U\Big]-t(\mu)
\Big(U^2-\dfrac{3}{2}\Big)\bigg\},
$$
$$
\lambda_C(\mu)=1-2\mu^2e^{-\mu^2}\int\limits_{0}^{1}e^{\mu\tau^2}d\tau,
$$
$$
t(\mu)=-2\mu e^{-\mu^2}\int\limits_{0}^{1}e^{\mu\tau^2}d\tau.
$$

We notice that
$$
q(-U,-U)=2\Big(U^2-\dfrac{1}{2}\Big)^2+2U^2+1>0,  \qquad
q(-U,0)=-\Big(U^2-\dfrac{3}{2}\Big).
$$

The function  $q(-U,\mu)$ has two real roots
$$
\mu_{1,2}^q=\dfrac{U\pm \sqrt{D(U)}}{2\Big(U^2-\dfrac{1}{2}\Big)},
$$
where
$$
D(U)=2\Big(U^2-\dfrac{3}{4}\Big)^2+\dfrac{3}{8}>0.
$$

We notice that
$$
\lim\limits_{U\to \pm \infty}\mu_1^q(U)=\dfrac{1}{\sqrt{2}},
$$

$$
\lim\limits_{U\to \pm \infty}\mu_2^q(U)=-\dfrac{1}{\sqrt{2}}.
$$

Let us consider family of curves
$\Gamma_U: z=G(t),-U \leqslant \mu \leqslant +\infty$.
These curves are closed: they begin and come to end in one point
$z=1$.

\begin{figure}[t]\center
\includegraphics[width=16.0cm, height=16cm]{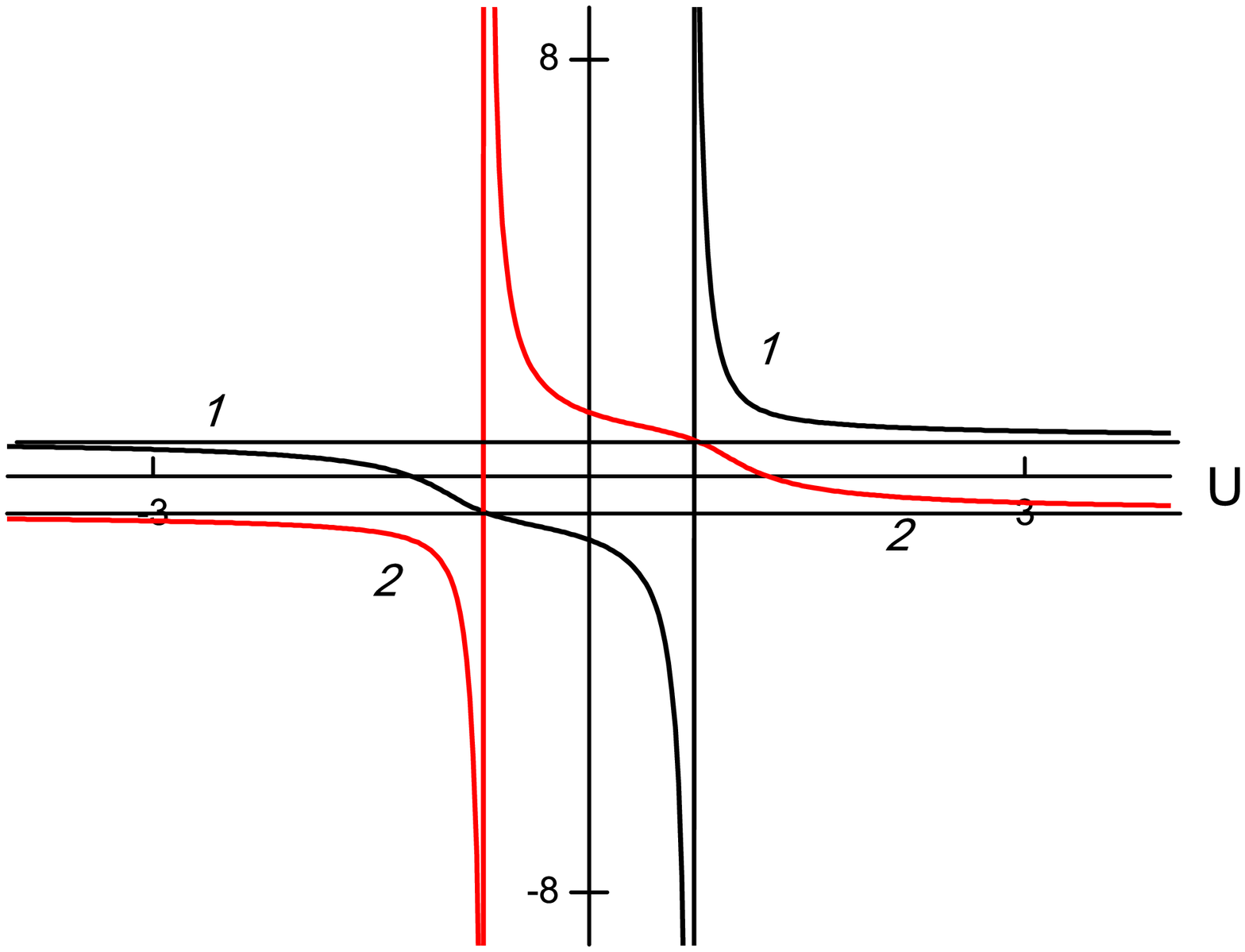}
{{\bf Fig. 5.}
Behavior of roots $\mu_1(U)$ (curve 1) and $\mu_2(U)$ (curve 2),
$U=\pm \dfrac{1}{\sqrt{2}}$ are vertical asymptotics,
$y=\pm \dfrac{1}{\sqrt{2}}$ are horisontal asymptotics.}
\end{figure}
\clearpage

 Really, it is easy to see, that
$$
\lambda(-U)=\lambda(+\infty)=1,\qquad s(-U)=s(+\infty)=0.
$$

For the solution of homogeneous Riemann problem (3.7)
impor\-tant calculate index of coefficient of problem $G (t) $
on the closed semiaxis $\mathbb{R}_U$, i.e. number of turns
concerning of origin of coordinates, which
the curve $\Gamma_U $ runs when the parametre $t $  describes
semiaxis $ \mathbb{R}_U $. As $ \arg G (t) =2\theta(t)$,
from here follows, that the number of turns of the curve
$\Gamma_U $ is equal to the doubled number of turns of the angle
$\theta(t)$ on the semiaxis
$ \mathbb{R}_U $, i.e. it is equal to the angle increment
$\theta(t)$ on the semiaxis $\mathbb{R}_U $, devided on $2 \pi $.
So for  finding angle increments $ \theta(t)$ we will consider
family of curves
$$
\gamma_U: z=\lambda^+(\mu)=\lambda(\mu)+is(\mu),\quad -U
\leqslant \mu \leqslant +\infty.
$$

At first we will consider the case of moderately strong evaporation
($U>0$).

Without the proof we will inform, that in the case $U> \sqrt {3/2} $
the angle increment $ \theta (t) $ on the semiaxis $ \mathbb{R}_U$
is equal $3 \pi $ (see Fig. 3). It means, the index of problem
is equal to three: $ \varkappa (G) =3\pi $.
In case of $0 <U <\sqrt {3/2} $ the angle increment
$ \theta (t) $ on the semiaxis $ \mathbb{R}_U $ is equal
$2 \pi $ (see Fig. 4). It means, the index of problem is equal to two:
$ \varkappa (G) =2\pi $.

Homogeneous Riemann problem (3.7) we will reduce to problem
of definition of analytical function on quantity of its jump on the cut
$$
\ln X^+(\mu)-\ln X^-(\mu)=2i[\theta(\mu)+k\pi],\quad -U\leqslant\mu
\leqslant+\infty,
\eqno{(3.8)}
$$
where $k=0,\pm 1,\pm 2, \cdots$.

In the case $U> \sqrt{3/2} $ in (3.8) it is necessary to take $k =-3$, i.e.
we consider the problem
$$
\ln X^+(\mu)-\ln X^-(\mu)=2i[\theta(\mu)-3\pi],\quad -U\leqslant\mu
\leqslant+\infty,
\eqno{(3.8')}
$$

As the solution  of problem (3.8') we take the limited in the point
$z =-U $ function
$$
X(z)=\dfrac{1}{(z+U)^3}e^{V(z)},
\eqno{(3.9)}
$$
where
$$
V(z)=\dfrac{1}{\pi}\int\limits_{-U}^{\infty}\dfrac{\theta(\mu)-3\pi}
{\mu-z}d\mu.
$$

By means of homogeneous Riemann boundary value problem
(3.7) we will transform
nonhomogeneous problem (3.6) to the problem of definition of the analytical
functions on its jump on the cut
$$
X^+(\mu)[q(-U,\mu)N^+(\mu)-h(0,\mu)]=$$$$=
X^-(\mu)[q(-U,\mu)N^-(\mu)-h(0,\mu)],\quad \mu>-U.
\eqno{(3.10)}
$$

Considering behaviour of the functions entering into
boundary value condition (3.10), we receive, that the problem
(3.10) has only trivial solution
$$
X(z)[q(-U,z)N(z)-h(0,z)] \equiv 0,
$$
whence we find
$$
N(z)=\dfrac{h(0,z)}{q(-U,z)}.
$$

However, this solution cannot be accepted as function $N(z)$,
entered by equality (3.5). This function is limited into
infinitely remote point, while the auxiliary
function (3.5) vanishes in infinitely remote point.

Thus, the considered boundary problem has no solution
in the case $U> \sqrt{3/2} $.

In case of $0 <U <\sqrt{3/2} $ in (3.8) it is necessary to take
$k =-2$, i.e. to consider the problem
$$
\ln X^+(\mu)-\ln X^-(\mu)=2i[\theta(\mu)-2\pi],\quad -U\leqslant\mu
\leqslant+\infty,
\eqno{(3.8'')}
$$

As the solution of problem (3.8") we take the limited in the point
$z =-U $ function
$$
X(z)=\dfrac{1}{(z+U)^2}e^{V(z)},
\eqno{(3.10)}
$$
where
$$
V(z)=\dfrac{1}{\pi}\int\limits_{-U}^{\infty}\dfrac{\theta(\mu)-2\pi}
{\mu-z}d\mu.
$$

Let us notice, that the angle $ \theta (\mu) $ (see Fig. 6)
conveniently to calculate under the formula
$$
\theta(\mu)=\arcctg\dfrac{\lambda(\mu)}{s(\mu)}+\left\{
\begin{array}{cc}
0 & -U\leqslant\mu\leqslant\mu_1^q(U), \\
\pi & \mu_1^q(U)\leqslant \mu \leqslant\mu_2^q(U), \\
2\pi & \mu_2^q(U)\leqslant\mu\leqslant+\infty.
\end{array}\right.
$$

\begin{figure}[h]\center
\includegraphics[width=16.0cm, height=10cm]{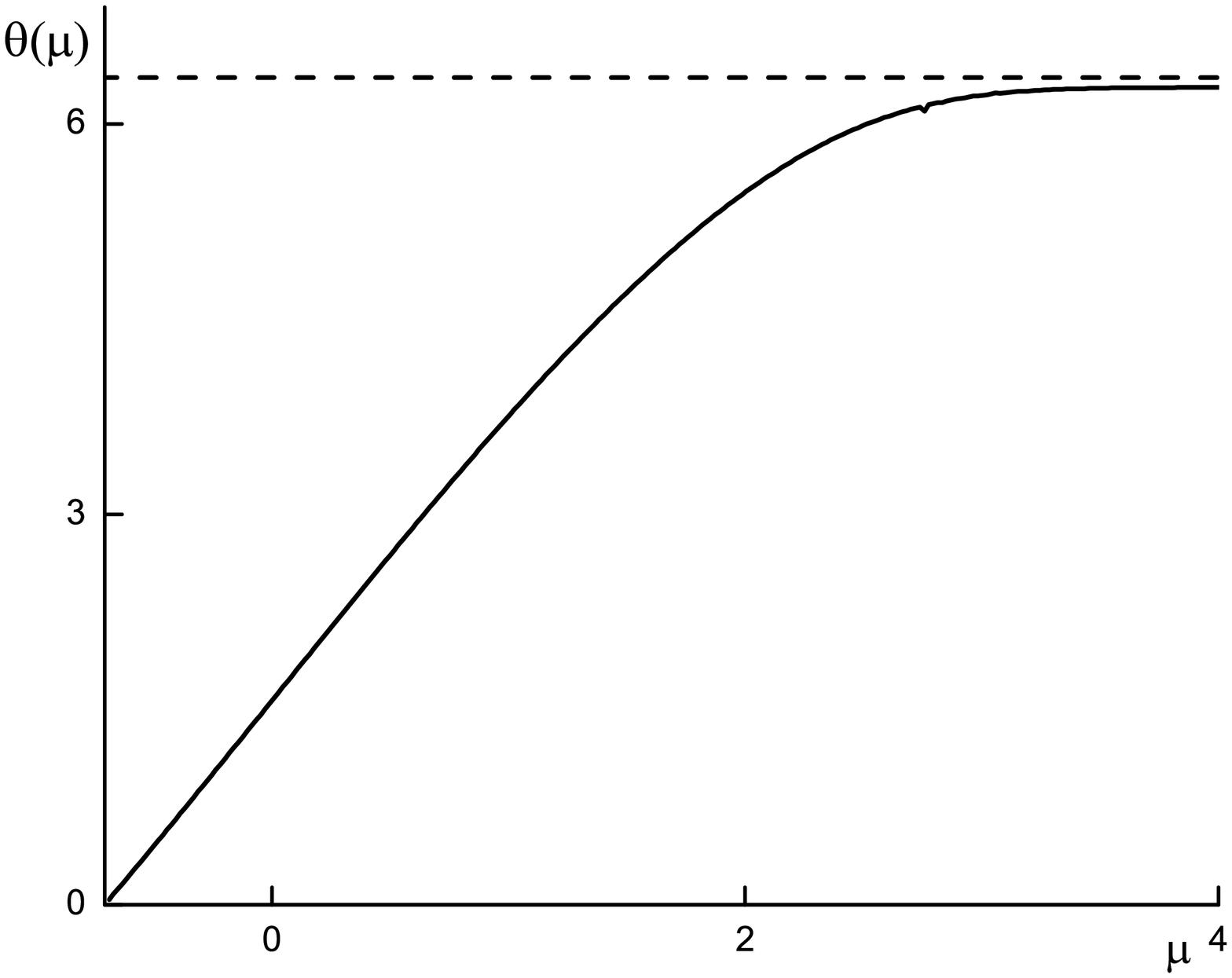}
{{\bf Fig. 6.} The angle $\theta(\mu)$ in case $U=1/\sqrt{2}$.
Incrementation of angle on  semiaxis $[-1/\sqrt{2},+\infty]$ equals $2\pi$.}
\end{figure}

In considered case the boundary value problem (3.10) has the solution
$$
X(z)[q(-U,z)N(z)-h(0,z)]=C_0,
$$
where $C_0$ is the arbitrary constant.

From this solution we find auxiliary function $N(z)$
$$
N(z)=\dfrac{h(0,z)+\dfrac{C_0}{X(z)}}{q(-U,z)}.
\eqno{(3.11)}
$$

The solution (3.11) represents meromorphic function.
Its denominator is function $q (-U, z) $. This function has in the case
$U\neq \dfrac{1}{\sqrt {2}} $
two real zero $ \mu_1^q (U) $ and $ \mu_2^q(U) $, and in the
case $U =\dfrac {1}{\sqrt{2}} $ function $q(-U, z) =1-\sqrt{2}z$
has unique zero. To this case we will return later.

Poles in points $ \mu_1^q(U) $ and $ \mu_2^q(U) $
are destroyed by conditions
$$
h(0,\mu_\alpha^q)+\dfrac{C_0}{X(\mu_\alpha^q)}=0,\quad
\alpha=1,2.
\eqno{(3.12)}
$$

Condition of vanishing of functions $N (z) $ in infinitely remote point
is reached by the condition
$$
C_0=-\varepsilon_T.
\eqno{(3.13)}
$$
Equality (3.13) follows from expansion in the Laurent series on
to negative degrees of $z$ the right part of equality (3.11).

Let us notice, that as function $N(z)$ is defined in the complex
plane, at its limiting values $N^{\pm}(\mu)$ from above and
from below in points $\mu_1^q(U) $ and $\mu_2^q(U)$ also exist
simple poles. That them to destroy, we will demand performance
four equalities
$$
h(0,\mu_\alpha^q)-\dfrac{\varepsilon_T}{X^{\pm}(\mu_\alpha)}=0,\quad
\alpha=1,2.
\eqno{(3.14)}
$$
Let us show, that these equalities coincide with equalities (3.12), i.e.
are carried out automatically. Really, we will result without a conclusion
integral representation of function $X(z)$
$$
X(z)=\dfrac{1}{\sqrt{\pi}}\int\limits_{-U}^{\infty}\dfrac{e^{-\mu^2}
(\mu+U)q(-U,\mu)X^+(\mu)}{\lambda^+(\mu)(\mu-z)}.
$$

From integral representation it is visible, that, as density
this integral in points $ \mu_1^q(U) $ and $ \mu_2^q(U) $ equals
to zero, boundary values of this integral from above and from below in
points $ \mu_1^q(U) $ and $ \mu_2^q(U) $ coincide with values
singular integral in these points. Thus,
$$
X^{\pm}(\mu_\alpha^q)=X(\mu_\alpha^q)+i\sqrt{\pi}
e^{-\mu^2}(\mu+U)q(-U,\mu_\alpha^q)\dfrac{X^+(\mu)}{\lambda^+(\mu)}=
X(\mu_\alpha^q),
$$
because $q(-U,\mu_\alpha^q)=0, \quad \alpha=1,2$.

From the equations (3.12) and (3.13) we find required
quantities of temperature jump and density jump as functions
of velocity of evapora\-tion $U$
$$
\varepsilon_T=2U\dfrac{(\mu_1^q-\mu_2^q)X(\mu_1^q)X(\mu_2^q)}
{({\mu_1^q}^2-{\mu_2^q}^2)X(\mu_1^q)X(\mu_2^q)+X(\mu_1^q)-X(\mu_2^q)}
\eqno{(3.15)}
$$
and
$$
\varepsilon_\rho=2U\mu_1^q-\Big[{\mu_1^q}^2-\dfrac{1}{2}-\dfrac{1}
{X(\mu_1^q)}\Big]\varepsilon_T.
\eqno{(3.16)}
$$

Let us return to the case $U =\dfrac {1}{\sqrt{2}} $.
The solution (3.11) let us present in the explicit form
$$
N(z)=\dfrac{C_0(z+U)e^{-V(z)}+\varepsilon_\rho-2Uz+\varepsilon_T
(z^2-1/2)}{1-\sqrt{2}z}.
$$

The pole is eliminated by one condition
$$
2C_0e^{-V(U)}+\varepsilon_\rho-1=0,
$$
whence we find
$$
C_0=\dfrac{1-\varepsilon_\rho}{2}e^{V(U)}.
$$

Destroying the pole of the second order in infinitely remote point,
we receive
$$
\varepsilon_T=-C_0=-\dfrac{1-\varepsilon_\rho}{2}e^{V(U)},
$$
and
$$
C_0(-V_1+\sqrt{2})-2U=0,
$$
where
$$
V_1=-\dfrac{1}{\pi}\int\limits_{-U}^{\infty}[\theta(\tau)-2\pi]d\tau.
$$
Expansion in vicinity infinitely remote point has been thus used
$$
e^{-V(z)}=1-\dfrac{V_1}{z}+\cdots,\qquad z\to \infty.
$$

From last equations we find quantities of jump of temperature and
density jump
$$
\varepsilon_T=-\dfrac{\sqrt{2}}{\sqrt{2}-V_1}=\dfrac{1}{V_1-\sqrt{2}}(2U),
$$
$$
\varepsilon_\rho=1-\dfrac{2\sqrt{2}e^{-V(U)}}{\sqrt{2}-V_1}=
-\dfrac{\sqrt{2}(1-2e^{-V(U)})-V_1}{\sqrt{2}(V_1-\sqrt{2})}(2U).
$$

Let us consider now various cases of condensation.

Let at first $-\sqrt {3/2} <U <0$. The analysis shows, that in
this case argument increment $ \theta(t) $ on the semiaxis
$ \mathbb{R}_U =[-U, + \infty] $ is equal $ \pi $.
Therefore as the solution of the homogeneous
Riemann problem we take limited in the point $z =-U $ function
$$
X(z)=\dfrac{1}{z+U}e^{V(z)},
$$
where
$$
V(z)=\dfrac{1}{\pi}\int\limits_{-U}^{\infty}\dfrac{\theta(\mu)-\pi}
{\mu-z}d\mu.
$$

The general solution of the problem (3.11) looks like now
$$
N(z)=\dfrac{\dfrac{C_0+C_1z}{X(z)}+h(0,z)}{q(-U,z)}.
\eqno{(3.17)}
$$

From the condition of vanishing of function $N(z)$ in infinitely remote
point, we find that
$$
C_1=-\varepsilon_T.
$$

From the condition of elimination of poles in points $ \mu_1^q $
and $ \mu_2$ we receive two equations
$$
X(\mu_\alpha)h(0,\mu_\alpha)+C_0+C_1\mu_\alpha=0,\qquad
\alpha=1,2.
$$

From this system we find
$$
C_0=\dfrac{1}{2}\Big\{\varepsilon_T\Big[\mu_1+\mu_2-X(\mu_1)
\Big(\mu_1^2-\dfrac{1}{2}\Big)-X(\mu_2)\Big(\mu_2^2-\dfrac{1}{2}\Big)\Big]+
$$$$+2U[\mu_1X(\mu_1)+\mu_2X(\mu_2)]-
\varepsilon_\rho[X(\mu_1)+X(\mu_2)]\Big\}
$$
and
$$
\varepsilon_T=2U\dfrac{\mu_1X(\mu_1)-\mu_2X(\mu_2)}{\mu_2-\mu_1+
X(\mu_1)(\mu_1-1/2)-X(\mu_2)(\mu_2-1/2)}
$$
$$
-\varepsilon_\rho\dfrac{X(\mu_1)-X(\mu_2)}{\mu_2-\mu_1+
X(\mu_1)(\mu_1-1/2)-X(\mu_2)(\mu_2-1/2)}.
$$

This solution is ambiguous. It contains the free unknown
parametre $\varepsilon_\rho$.

Let now $U <-\sqrt{3/2} $. In this case it is possible to show, that
increment of angle $ \theta(\mu)$ on the semiaxis $ \mathbb{R}_U =
[-U, + \infty] $ is equal to zero.
Therefore as the solution of the homogeneous
Riemann problem we take limited in the point $z =-U $ function
$$
X(z)=e^{V(z)},
$$
where
$$
V(z)=\dfrac{1}{\pi}\int\limits_{-U}^{\infty}\dfrac{\theta(\mu)d\mu}
{\mu-z}.
$$

Hence, the general solution of a nonhomogeneous boundary value
problem (3.10) looks like
$$
N(z)=\dfrac{\dfrac{C_0+C_1z+C_2z^2}{X(z)}+h(0,z)}{q(-U,z)},
$$
where $C_0, C_1, C_2$ are arbitrary constans.

From  condition of vanishing of solution in infinitely remote point
we find
$$
C_2=-\varepsilon_T.
$$

From condition of elimination of poles we have the following equalities
$$
C_0+C_1\mu_\alpha^q+C_2{\mu_\alpha^q}^2+
h(0,\mu_\alpha^q){X(\mu_\alpha^q)}=0,\quad \alpha=1,2.
$$

From this system we find
$$
C_0=\varepsilon_T(\mu_1^q+\mu_2^q)-\dfrac{X(\mu_1)h(0,\mu_1^q)-
X(\mu_2^q)h(0,\mu_2^q)}{\mu_1^q-\mu_2^q}
$$
and
$$
C_0=\dfrac{1}{2}\Big\{-C_1(\mu_1^q+\mu_2^q)+\varepsilon_T
({\mu_1^q}^2+{\mu_2^q}^2)-X(\mu_1^q)h(0,\mu_1^q)-
X(\mu_1^q)h(0,\mu_2^q)\Big\}.
$$

This solution, as well as previous, ambiguously. It contains two
free parametres $\varepsilon_T$ и $\varepsilon_\rho$.

For the unequivocal solution of the problem on condensation it is necessary
to set three parametres
$U$, $\varepsilon_T$ и $\varepsilon_\rho$.

\begin{center}
  {\bf 4.  Temperature jump and weak evaporation (condensation).
Distribution of  gas macroparameters}
\end{center}

Quantities of jumps of temperature and density we will present in
the following form
$$
\varepsilon_T=K(U)(2U),\qquad \varepsilon_\rho=R(U)(2U),
$$
where coefficients of jumps of temperature and density $K(U)$ and
$R(U)$ are given by formulas
$$
K(U)=\dfrac{(\mu_1^q-\mu_2^q)X(\mu_1^q)X(\mu_2^q)}
{({\mu_1^q}^2-{\mu_2^q}^2)X(\mu_1^q)X(\mu_2^q)+X(\mu_1^q)-X(\mu_2^q)}
$$
and
$$
R(U)=\mu_1^q-\Big[{\mu_1^q}^2-\dfrac{1}{2}-\dfrac{1}
{X(\mu_1^q)}\Big]K(U).
$$

\begin{figure}[h]\center
\includegraphics[width=16.0cm, height=8cm]{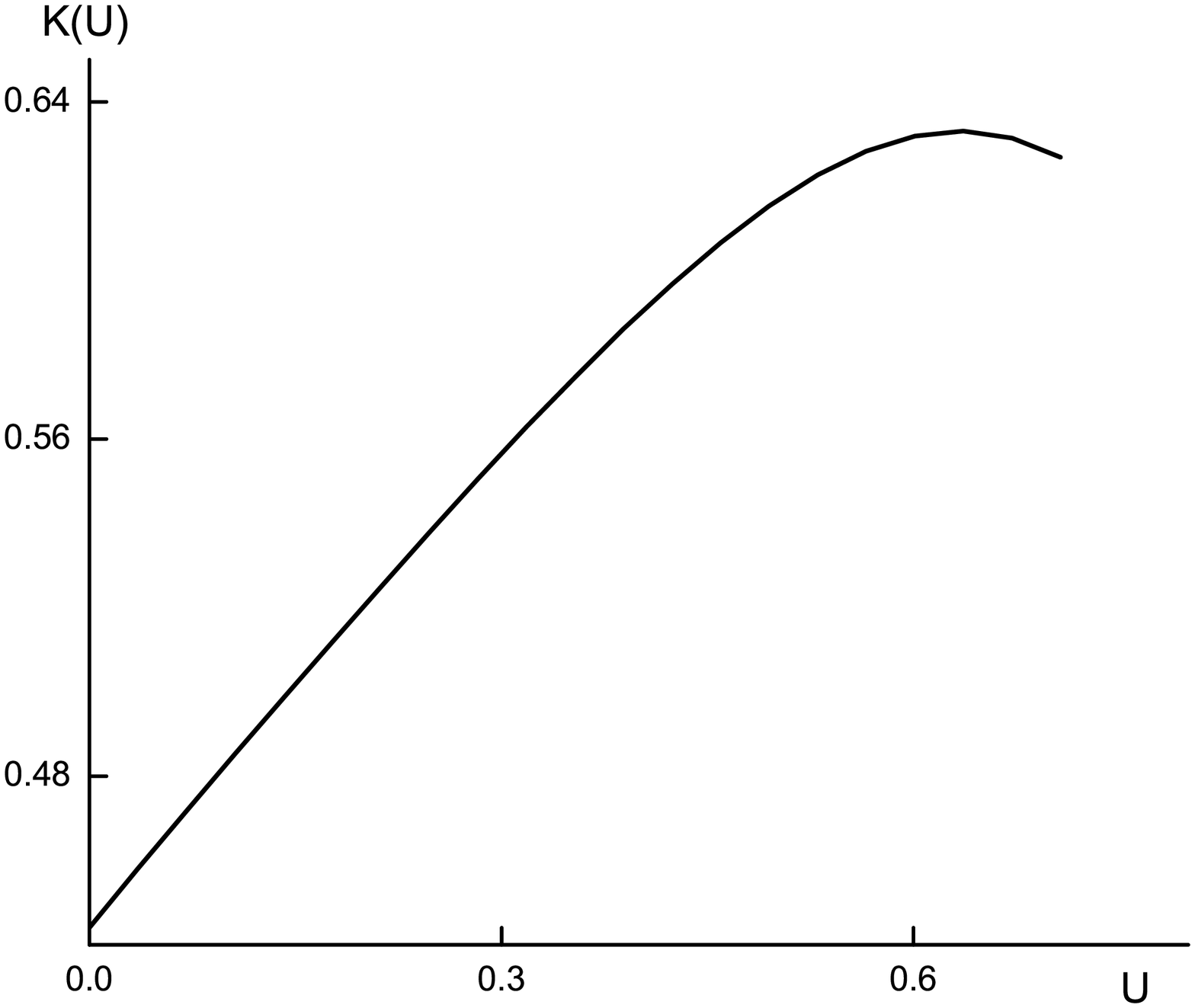}
{{\bf Fig. 7.} Dependence of temperature jump coefficient on
dimensionless quantity of evaporation velocity in the range
$0\leqslant U \leqslant 1/\sqrt{2}$.}
\end{figure}
\begin{figure}[h]\center
\includegraphics[width=16.0cm, height=8cm]{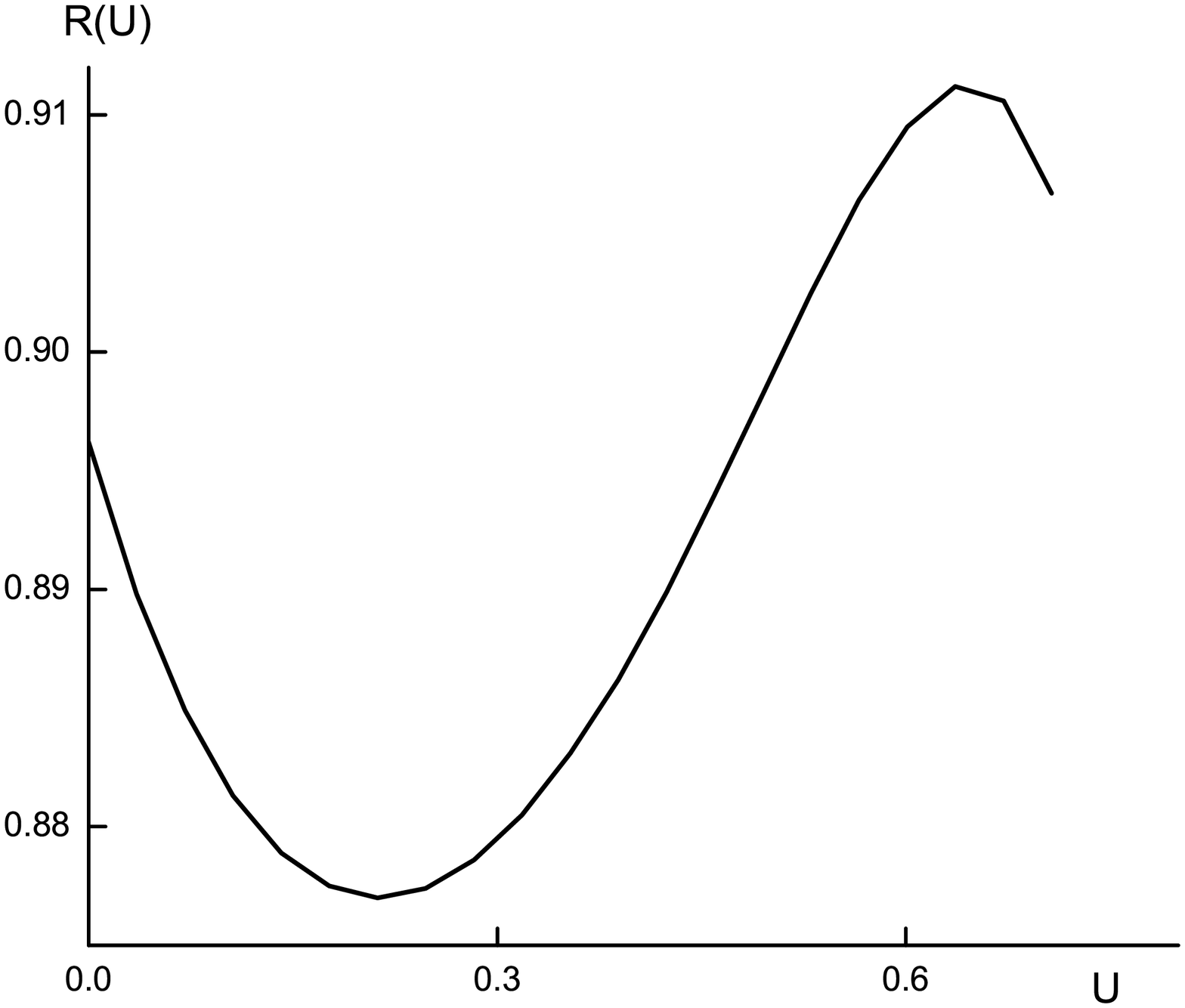}
{{\bf Fig. 8.} Dependence of density jump coefficient on
dimensionless quantity of evaporation velocity in the range
$0\leqslant U\leqslant 1/\sqrt{2}$.}
\end{figure}
\clearpage

Let us carry out numerical calculations in the case
$U=\dfrac{1}{\sqrt{2}}$. In this case $V_1=3.0095, V(U)=1.8376$.
Therefore
$$
\varepsilon_T=0.6268(2U_0)=0.8864, \quad
\varepsilon_\rho=-1.7612(2U_0)=-2.4907.
$$

For comparison we will present results on weak evaporation with
application of the one-dimensional kinetic equation with
collisional fre\-qu\-ency, proportional to the module
of molecular velocity
$$
\varepsilon_T=-0.5046(2U),\qquad
\varepsilon_n=-0.2523(2U).
$$

For comparison we will present coefficients of jump of temperature and jump
of concentration found by means of the one-dimensional kinetic
equations with constant frequency of collisions \cite{10}
$$
\varepsilon_T=-0.4443(2U),\qquad
\varepsilon_n=-0.8958(2U).
$$

Coefficients of continuous spectrum we will find from
Sokhotsky' formulas  for auxiliary function and the constructed
solution
$$
N^+(\mu)-N^-(\mu)=2\sqrt{\pi}i(\mu+U)a(\mu)=
-\dfrac{\varepsilon_T}{q(-U,\mu)}\Big[\dfrac{1}{X^+(\mu)}-
\dfrac{1}{X^-(\mu)}\Big].
$$
It is easy to see, that
$$
\dfrac{1}{X^+(\mu)}-\dfrac{1}{X^-(\mu)}=-2i\dfrac{\sin
\theta(\mu)}{X(\mu)}.
$$
Hence, the factor of the continuous spectrum is equal
$$
(\mu+U)a(\mu)=\dfrac{\sin \theta(\mu)}{\sqrt{\pi}q(-U,\mu)X(\mu)}
\varepsilon_T.
\eqno{(4.1)}
$$

Substituting the found expression in expansion (3.2), we find
dis\-con\-ti\-nuous in the point $ \mu =-U $ at $x=0$ distribution function
$$
\dfrac{h(x,\mu)}{\varepsilon_T}=\dfrac{1}{\sqrt{\pi}}
\int\limits_{-U}^{\infty}\exp\Big(-\dfrac{x}{\eta+U}\Big)
\dfrac{\sin\theta(\eta)}{X(\eta)}d\eta+
$$
$$
+\exp\Big(\mu^2-\dfrac{x}{\mu+U}\Big)
\dfrac{\lambda(\mu)\sin\theta(\mu)}
{\sqrt{\pi}q(-U,\mu)X(\mu)}\Theta_+(\mu+U).
$$

Here $\Theta_+(x)$ is the known Heaviside function (unit jump at
origin of coordinates).

Substituting coefficient of the continuous spectrum (4.1) in
distribution of gas density (1.3), and considering, that
normalization of eigen function of zero order is identically
equal to unit, we receive
$$
\dfrac{\rho(x)}{\rho_\infty}=1+\dfrac{1}{\sqrt{\pi}}
\int\limits_{-U}^{\infty}e^{-x/(\eta+U)}d\eta\int\limits_{-\infty}^{\infty}
e^{-\mu^2}\Phi(\eta,\mu)d\mu=
$$
$$
=1+\dfrac{\varepsilon_T}{\pi}
\int\limits_{-U}^{\infty}e^{-x/(\eta+U)}\dfrac{\sin \theta(\eta)d\eta}
{(\eta+U)q(-U,\eta)X(\eta)}.
$$

Let us substitute (4.1) in distribution of mass velocity
(1.4) and considering, that normalization of the first order
is equal $-U $, we receive
$$
U(x)=U+\dfrac{1}{\sqrt{\pi}}
\int\limits_{-U}^{\infty}e^{-x/(\eta+U)}d\eta\int\limits_{-\infty}^{\infty}
e^{-\mu^2}\mu\Phi(\eta,\mu)d\mu=
$$
$$
=U\Bigg[1-\dfrac{\varepsilon_T}{\pi}
\int\limits_{-U}^{\infty}e^{-x/(\eta+U)}\dfrac{\sin \theta(\eta)d\eta}
{(\eta+U)q(-U,\eta)X(\eta)}\Bigg].
$$

Substituting (4.1) in distribution of temperature (1.5) and considering,
that normalization of the second order is equal $U^2$, we receive
$$
\dfrac{T(x)}{T_\infty}=1+\dfrac{2}{\sqrt{\pi}}
\int\limits_{-U}^{\infty}e^{-x/(\eta+U)}a(\eta)d\eta
\int\limits_{-\infty}^{\infty}e^{-\mu^2}
\Big(\mu^2-\dfrac{1}{2}\Big)\Phi(\eta,\mu)d\mu=
$$
$$
=1+2\Big(U^2-\dfrac{1}{2}\Big)\dfrac{\varepsilon_T}{\pi}
\int\limits_{-U}^{\infty}e^{-x/(\eta+U)}\dfrac{\sin \theta(\eta)d\eta}
{(\eta+U)q(-U,\eta)X(\eta)}.
$$

From last formulas it is visible, that distribution of mass
velocity and density are connected by equality
$$
\dfrac{\rho(x)}{\rho_\infty}+\dfrac{U(x)}{U_\infty}=2,
$$
and density and temperature distribution are connected by equality
$$
\dfrac{T(x)}{T_\infty}=1+2\Big(U^2-\dfrac{1}{2}\Big)\Big[
\dfrac{\rho(x)}{\rho_\infty}-1\Big].
$$
\begin{center}
  \bf 5. Conclusion
\end{center}

In the present work the analytical solution of
boundary problem about moderately strong evaporation (condensation)
with application of the one-dimensional kinetic equation with
constant frequency of collisions of molecules is considered.

The carried out analysis shows, that both with physical, and with
mathematical point of view of a problem of evaporation and condensation
are non-symmetrical.

Let us notice, that for one-dimensional gas value
of the Mach number $ {\bf M} =1$
corresponds to velocity of evaporation (condensation) $U =\sqrt{3/2} $.

From results of work follows, that this quantity renders
essential influence on evaporation and condensation modes.

The modes of condensation established in work at the various
values of velocity of condensation $U <0$ (subsonic and supersonic
cases) correspond to results of numerical calculations,
done in work \cite{7}.

Function of distribution of gas molecules in explicit form
is received, and also distributions of density of gas,
its mass velocity and temperature in half-space $x> 0$.

\end{document}